\documentclass[a4paper]{jpconf}
\usepackage{graphicx}
\newcommand{\be}{\begin{equation}}
\newcommand{\bea}{\begin{eqnarray}}
\newcommand{\ee}{\end{equation}}
\newcommand{\eea}{\end{eqnarray}}
\newcommand{\bpi}{\begin{picture}}
\newcommand{\bce}{\begin{center}}

\def\gv{\widetilde{{\rm I}\hspace{-0.07cm}\Gamma}}

\begin{document}
\title{Gluon mass and freezing of the QCD coupling}

\author{ A.C. Aguilar$^a$ and J. Papavassiliou$^a$}

\address{$^a$ Departamento de F\`\i sica Te\`orica and IFIC
Centro Mixto, Universidad de Valencia-CSIC, E-46100, Burjassot,
Valencia, Spain }

\ead{joannis.papavassiliou@uv.es}

\begin{abstract}

Infrared finite  solutions for  the gluon propagator  of pure  QCD are
obtained from the gauge-invariant non-linear Schwinger-Dyson equation
formulated in the Feynman gauge of the background field method.  These
solutions may be  fitted using a massive propagator,  with the special
characteristic   that   the    effective   ``mass''   employed   drops
asymptotically  as the  inverse square  of the  momentum  transfer, in
agreement with  general operator-product expansion  arguments.  Due to
the presence of  the dynamical gluon mass the  strong effective charge
extracted from these solutions freezes  at a finite value, giving rise
to an infrared fixed point for QCD.

\end{abstract}

\section{Introduction}

The systematic study of Schwinger-Dyson equations (SDE) 
in the framework of the pinch technique (PT) has led to the conclusion that
the non-perturbative QCD dynamics generate 
an effective, mometum-dependent mass for the gluon, while  preserving 
the   local  $SU(3)_c$   invariance 
of the theory~\cite{Cornwall:1981zr,Aguilar:2006gr,Aguilar:2007ie}.
This picture is further corroborated 
by lattice simulation and a 
variety of theoretical and phenomenological works~\cite{ref}.
One of the most important consequences of this 
picture is that this dynamical mass
tames  the   Landau  singularity   associated   with  the
perturbative $\beta$  function, giving rise to a
strong effective charge ``freezing'' at a  finite value in the infrared.
In this talk we report recent progress in the study of a 
non-linear SDE for the gluon propagator~\cite{Aguilar:2007ie}.

\section{The non-linear SDE}  
\begin{figure}[h]
\hspace{2.0cm}
{\includegraphics[scale=1.0]{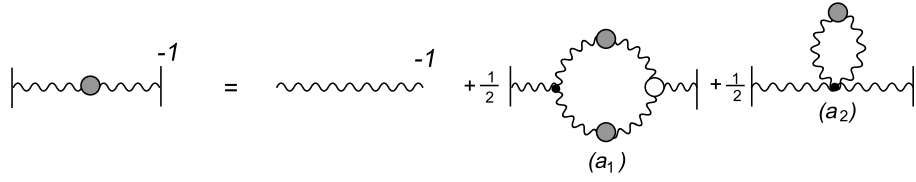}}
\caption{The gluonic ``one-loop dressed'' contributions to the SDE.}
\label{f1}
\end{figure}

The relevant SDE for $\Delta_{\mu\nu}(q)$ is shown in Fig.(\ref{f1}). Due to the special 
properties of the truncation scheme based on the PT~\cite{Cornwall:1981zr,Cornwall:1989gv}(and its connection
with the Feynman gauge of the background field method (BFM)~\cite{Abbott:1980hw}), 
this equation is gauge-invariant despite 
the omission of ghost loops or higher order graphs~\cite{Aguilar:2006gr}.
Dropping for simplicity the longitudinal momenta, i.e. setting $\Delta_{\mu\nu}(q)= -i g_{\mu\nu}\Delta(q^2)$,
one looks for solutions where  $\Delta(q^2)$ reaches a finite
(non-vanishing) value in the  deep infrared;  
such solutions may
be  fitted  by     ``massive''  propagators  of   the 
form $\Delta^{-1}(q^2)  =  q^2  +  m^2(q^2)$,  
where $m^2(q^2)$ is  not ``hard'', but
depends non-trivially  on the momentum  transfer $q^2$.  
The tree-level expressions for the three- and four-gluon vertices appearing in the two graphs
of Fig.(\ref{f1}) are given in the first item of~\cite{Abbott:1980hw}.
For the full three-gluon vertex, $\gv$, denoted by the white blob in graph $({a_1})$, we 
employ a gauge technique Ansatz, expressing 
it as a functional of $\Delta$, 
in such a way as to satisfy (by construction) the all-order Ward identity 
\begin{equation}
q^{\mu}\gv_{\mu\alpha\beta}(q,p_1,p_2) = 
i[{\Delta}^{-1}_{\alpha\beta}(p_1) - {\Delta}^{-1}_{\alpha\beta}(p_2)]\,,
\label{ward1}
\end{equation}
characteristic of the PT-BFM.
Specifically, we use the following closed form for the vertex~\cite{Aguilar:2007ie}:
\bea
\gv^{\mu\alpha\beta} &=& \widetilde{\Gamma}^{\mu\alpha\beta}
+i g^{\alpha\beta}\, \frac{q^{\mu}}{q^2}\,
\left[ {\Pi}(p_2) - {\Pi}(p_1)\right]
- i\frac{c_1}{q^2}\left(q^{\beta}g^{\mu\alpha} - q^{\alpha}g^{\mu\beta}\right)
\left[{\Pi}(p_1)+ {\Pi}(p_2)\right]\nonumber\\
&&-i c_2 \left(q^{\beta}g^{\mu\alpha} - q^{\alpha}g^{\mu\beta}\right)
\left[\frac{{\Pi}(p_1)}{p_1^2}+ \frac{{\Pi}(p_2)}{p_2^2}\right]\,.
\label{gtvertex}
\eea
with $\widetilde{\Gamma}_{\mu\alpha\beta}(q,p_1,p_2)=  
(p_1-p_2)_{\mu} g_{\alpha\beta} + 2\,q_{\beta}g_{\mu\alpha} - 2\,q_{\alpha}g_{\mu\beta}$,
and $i\Pi(q^2) =  \Delta^{-1}(q^2) - q^2$.

\vspace{0.2cm} 

Defining the renormalization-group invariant quantity~\cite{Cornwall:1989gv} ${d}(q^2) = g^2 \Delta(q^2)$, we arrive at
\be
d^{-1}(x) = K^{\prime}x + \tilde{b} \sum_{i=1}^8 {\widehat A}_i(x)\, + d^{-1}(0)\,,
\label{rgisde}
\ee
with
\bea
{\widehat A}_1(x) &=& - \left(1+\frac{6 c_2}{5}\right) x \int_{x}^{\infty}\!\!\! dy \,y\, {\cal L}^{\,2}(y) d^{\,2}(y)\,,   
\nonumber\\ 
{\widehat A}_2(x) &=& \frac{6 c_2}{5} x \int_{x}^{\infty}\!\!\! dy\, {\cal L}(y) d(y)\,, 
\nonumber\\
{\widehat A}_3(x) &=& -\left(1+\frac{6 c_2}{5}  \right) x\, {\cal L}(x) d(x)\int_{0}^{x}\!\!\! dy\, y\, {\cal L}(y) d(y)\,, 
\nonumber\\ 
{\widehat A}_4(x) &=& \left(-\frac{1}{10} -  \frac{3c_2}{5} + \frac{3c_1}{5}\right)
\int_{0}^{x}\!\!\! dy\, y^2\, {\cal L}^{\,2}(y) d^{\,2}(y)\,, 
\nonumber \\
{\widehat A}_5(x) &=& - \frac{6}{5}\bigg(1+ c_1\bigg)  {\cal L}(x) d(x)\int_{0}^{x}\!\!\! dy \,y^2\,{\cal L}(y) d(y)\,, 
\nonumber \\
{\widehat A}_6(x) &=& \frac{6c_2 }{5} \int_{0}^{x}\!\!\! dy\, y \,{\cal L}(y) d(y)\,,
\nonumber\\
{\widehat A}_7(x) &=& \frac{2}{5}\, {\cal L}(x)\,\frac{d(x)}{x}\int_{0}^{x}\!\!\! dy\, y^3\,  {\cal L}(y) d(y)\,, 
\nonumber\\
{\widehat A}_8(x) &=& \frac{1}{5x} \int_{0}^{x} \!\!\! dy\, y^3\,  {\cal L}^{\,2}(y) d^{\,2}(y)\,,
\label{hatAi}
\eea
where $x=q^2$.
The renormalization constant $ K^{\prime}$ is fixed by the condition $d^{-1}(\mu^2)=\mu^2/g^2$, (with 
$\mu^2\gg \Lambda^2$), and 
${\cal L}(q^2) \equiv \tilde{b}\ln\left(q^2/\Lambda^2\right)$, 
where $\Lambda$ is QCD mass scale.
Due to the poles contained in the Ansatz for $\gv^{\mu\alpha\beta}$, $d^{\,-1}(0)$ does not vanish, 
and is given by the (divergent) expression
\be
d^{\,-1}(0) = \frac{3 \tilde{b}}{5\pi^2} 
\Bigg[2(1+c_1)\int  d^4 k \,{\cal L}(k^2)\, d(k^2) - (1+2c_1)\int\,d^4 k \, k^2 \,{\cal L}^{\,2}(k^2)\,d^{\,2}(k^2) \Bigg]\,,
\label{D0}
\ee
which can be made finite using dimensional regularization, and assuming 
that $m^2(q^2)$ drops sufficiently fast in the UV~\cite{Aguilar:2006gr}.

\section{Results}

The way to extract from  $d(q^2)$ the corresponding $m^2(q^2)$ and $g^2(q^2)$
is by casting the numerical solutions into the form~\cite{Cornwall:1981zr} 
\be
d(q^2) = \frac{g^2(q^2)}{q^2+m^2(q^2)} \,, \quad
g^2(q^2) = \bigg[ \tilde{b}\ln\left(\frac{q^2 + f(q^2, m^2(q^2))}{\Lambda^2}\right)\bigg]^{-1}\,.
\label{GNP}
\ee
with
\be
f(q^2, m^2(q^2)) = \rho_{\,1} m^2(q^2)+ \rho_{\,2} \frac{m^4(q^2)}{q^2+m^2(q^2)} 
+\rho_{\,3} \frac{m^6(q^2)}{[q^2+m^2(q^2)]^{\,2}} \,,
\label{func_fit}
\ee
The functional form  used for the running mass is
\be
m^2(q^2)=\frac{m^4_0}{q^2+m^2_0}\Bigg[\ln
\left(\frac{q^2+\rho\,m^2_0}{\Lambda^2}\right)\Big/\ln\left(\frac{\rho\,m^2_0}{\Lambda^2}\right) \Bigg]^{\gamma_2-1} \,,
\label{dmass_power}
\ee 
where $\gamma_2= \frac{4}{5} + \frac{6c_1}{5}$; $\rho$, $\rho_1$, $\rho_2$, and  $\rho_3$
are adjustable constants.
Evidently, $m^2(q^2)$ is   
dropping in the deep ultraviolet as an {\it inverse power} of the momentum,
as expected from general operator-product expansion 
calculations~\cite{Lavelle:1991ve}.
Note that $ f(q^2, m^2(q^2))$ is such that $ f(0, m^2(0)) >0$; as a result, 
$g^2(q^2)$ reaches a  finite positive value  at $q^2=0$,
leading to an {\it infrared fixed point}
\cite{Cornwall:1981zr,Aguilar:2002tc,Brodsky:2007wi}.

\vspace{0.5cm}
\begin{figure}[ht]
\hspace{.5cm}
\includegraphics[scale=2.0]{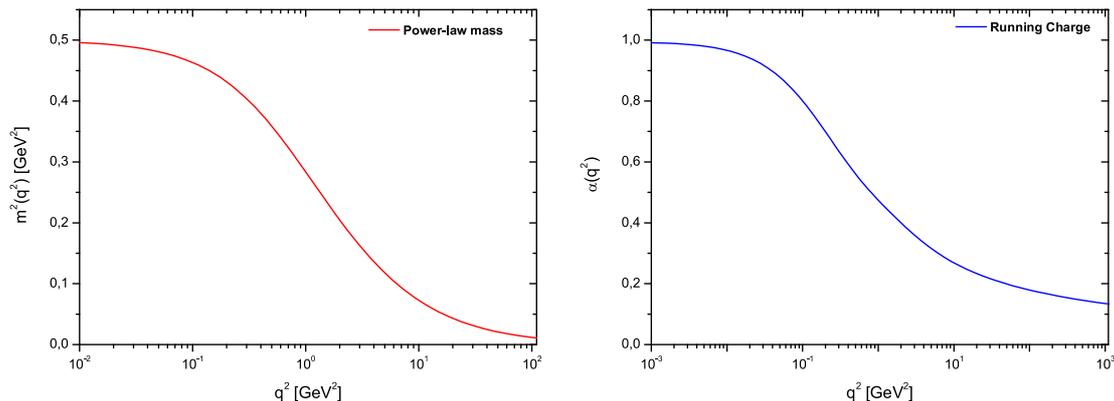}
\vspace{-0.5cm}
\caption{\small{
Left: dynamical mass with power-law running, for 
$m_0^2=0.5    \;\mbox{GeV}^{\,2}$     and    $\rho=1.046$    in  
Eq.(\ref{dmass_power}). Right: the running charge,
$\alpha(q^2)=g^2(q^2)/4\pi$.}}
\end{figure}

\subsection{Acknowledgments}
This work was supported by the Spanish MEC under the grants FPA 2005-01678
and FPA 2005-00711, and the Fundaci\'on General of the University of Valencia.

\end{document}